\documentclass{article}

\pdfoutput=1

\usepackage{amsmath}
\usepackage{amsfonts}
\usepackage{amssymb}
\usepackage{amsthm}
\usepackage[numbers,sort&compress]{natbib}
\usepackage{bm}
\usepackage{graphicx}
\usepackage{subfigure}
\usepackage{array}
\usepackage{multirow}
\usepackage{algorithm}
\usepackage{algorithmic}
\usepackage{color}
\usepackage[normalem]{ulem}
\usepackage{authblk}

\DeclareMathAlphabet{\mathsfsl}{OT1}{cmss}{m}{sl}
\newcommand{\tensor}[1]{\bm{#1}}

\newcommand{\bx}{\bm\xi}

\newcommand{\mean}[1]{\left\langle {#1} \right\rangle}

\newcommand{\PreserveBackslash}[1]{\let\temp=\\#1\let\\=\temp}
\newcolumntype{C}[1]{>{\PreserveBackslash\centering}p{#1}}
\newcolumntype{L}[1]{>{\PreserveBackslash\raggedright}p{#1}}

\newcommand{\revision}[1]{{#1}}

\author[1,*]{Xiu Yang}
\author[1,*]{Huan Lei}
\author[1]{Peiyuan Gao}
\author[2]{Dennis G.~Thomas}
\author[3]{David Mobley}
\author[1,4,5]{Nathan A.~Baker}

\affil[*]{These authors contributed equally.}
\affil[1]{Advanced Computing, Mathematics, and Data Division, Pacific Northwest National Laboratory, Richland, WA 99352, USA}
\affil[2]{Biological Sciences Division, Pacific Northwest National Laboratory, Richland, WA 99352, USA}
\affil[3]{Department of Pharmaceutical Sciences, University of California Irvine, Irvine, CA 92697, USA}
\affil[4]{Division of Applied Mathematics, Brown University, Providence, RI 02912, USA}
\affil[5]{Contact information:  \texttt{nathan.baker@pnnl.gov}, {+1-509-375-3997}}

\title{Atomic radius and charge parameter uncertainty in biomolecular solvation energy calculations}

\begin{document}

\maketitle

\begin{abstract}
	Atomic radii and charges are two major parameters used in implicit solvent electrostatics and energy calculations.
	The optimization problem for charges and radii is under-determined, leading to uncertainty in the values of these parameters and in the results of solvation energy calculations using these parameters.
	This paper presents a \revision{new method for quantifying this uncertainty in implicit solvation calculations of small molecules} using surrogate models based on generalized polynomial chaos (gPC) expansions.
	There are relatively few atom types used to specify radii parameters in implicit solvation calculations; therefore, surrogate models for these low-dimensional spaces could be constructed using least-squares fitting.
	However, there are many more types of atomic charges; therefore, construction of surrogate models for the charge parameter space requires compressed sensing combined with an iterative rotation method to enhance problem sparsity.
	\revision{We demonstrate the application of the method by presenting results for the uncertainties in small molecule solvation energies based on these approaches.}
	The method presented in this paper is a promising approach for efficiently quantifying uncertainty in a wide range of force field parameterization problems, including those beyond continuum solvation calculations.
	\revision{The intent of this study is to provide a way for developers of implicit solvent model parameter sets to understand the sensitivity of their target properties (solvation energy) on underlying choices for solute radius and charge parameters.}
\end{abstract}

\section{Introduction}
Implicit solvent models and their applications have been the subject of numerous previous reviews \cite{Lamm03, Ren2012, GroTry08}.
Such solvation models require the coordinates of the solute atoms as well as atomic charge distributions and a representation of the solute-solvent interface.
Charges and interfaces are generally modeled through parameterized empirical representations; however, these parameterizations are often under-determined, leading to uncertainty in the resulting parameter sets \cite{PoCa03, Gosink16, Swanson05}.
\revision{The Poisson equation is a popular model for implicit solvent electrostatics and serves as a good example for exploring the influence of this uncertainty on properties such as molecular solvation energy \cite{Lamm03, Ren2012, GroTry08, LiLPA13}.
In this paper, we use the term ``solvation energy'' to refer to the energy returned from the Poisson equation and to emphasize that we are not sampling over solute conformational states for a true \emph{free} energy.}
This is a partial differential equation for the electrostatic potential $\varphi : \Omega \mapsto \mathbb{R}$
\begin{align}
-\nabla \cdot \epsilon(\bm x) \nabla \varphi(\bm x) &= \rho(\bm x) \text{~for~} \bm x \in \Omega
\label{eqn:poisson} \\
\varphi(\bm s) &= \varphi_D(\bm s) \text{~for~} \bm s \in \partial \Omega,
\end{align}
where $\Omega \subset \mathbb{R}^3$ is the problem domain, $\partial \Omega$ is
the domain boundary, $\epsilon : \Omega \mapsto [1, \infty)$ is a dielectric
coefficient, $\rho: \Omega \mapsto \mathbb{R}$ is the charge distribution, and
$\varphi_D$ is a reference potential function (e.g., Coulomb's law) used for the Dirichlet boundary condition.
The dielectric coefficient $\epsilon$ is usually defined implicitly \cite{Swanson05b, Swanson07, Bates08, Dong06} with respect to the solute atomic radii $\{ \sigma_i \}$ and solvent properties such that the coefficient reaches two limiting constant values:  $\epsilon_u$ inside the solute and $\epsilon_v$ away from the solute in bulk solvent.
The solvation energy is calculated by
\begin{equation}
	\Delta G = \int_{\Omega} \rho(\bm x) \left( \varphi(\bm x) - \varphi_0(\bm x) \right)d \bm x,
	\label{eqn:solvation-generic}
\end{equation}
where $\varphi$ is the Poisson equation solution for the system with a bulk value of $\epsilon$ corresponding to the solvent of interest and $\varphi_0$ is the solution for the system with a bulk value of $\epsilon$ corresponding to a vacuum.
For atomic monopoles, the solute charge distribution has the (numerically unfortunate) form
$\rho(\bm x) = \sum_i^{N_A} q_i \delta(\bm x - \bm x_i)$
for $N_A$ solute atoms with positions $\{\bm x_i\}$ and charges $q_i$.
The $\delta$ terms are formally defined as Dirac delta functionals but usually approximated by functions with finite support (e.g., when projected onto a grid or finite element basis).
The delta functional approximation leads to a simplified form for the solvation energy in Eq.~\ref{eqn:solvation-generic},
\begin{equation}
	\Delta G = \sum_i^{N_A} q_i \left( \varphi(\bm x_i) - \varphi_0(\bm x_i) \right).
\end{equation}

\revision{Charges and interfaces in implicit solvent representations are generally modeled through
parameterized empirical representations; however, these parameterizations are
often under-determined leading to uncertainty in the resulting parameter sets \cite{PoCa03, Gosink16, Swanson05}.
For example, atomic} charge models are designed to approximate the ``true'' vacuum  electrostatic potential due to quantum mechanical electron and nuclei charge distributions.
While quantum mechanical charge distributions can be incorporated directly in implicit solvent models \cite{Eckert2010, Tomasi2005}, atomic point charge distributions are generally used \cite{Ren2012}.
These point charges can include inducible and fixed multipoles \cite{GK, SchPB} but monopoles are the most common form.
For the purposes of assigning charges, atoms are grouped into sets based on molecular connectivity and environment \cite{NichollsMGCBCP08}.
The charge values for atoms in these sets are usually determined by numerical fitting to quantum mechanical vacuum electrostatic potentials.
Such charge optimization is ill-posed and fitting requires careful choice of the objective function and regularization constraints \cite{BeslerMK90, HaschkaHJMED15, resp, AIM}.
While  sophisticated fitting procedures have been developed, significant information reduction occurs in the transformation of the continuous quantum mechanical electron density into a discrete set of atomic point charges.

Solute-solvent interface models are much more empirical than the charge distribution models; the definition of a solvent ``interface'' is imprecise at length scales comparable to the size of water molecules.
Therefore, such models are generally developed to represent a reasonable description of the solute geometry while also optimizing agreement with experimental quantities such as solvation energy.
A large number of solute-solvent interface models exist, including van der Waals
\cite{Dong06}, solvent-accessible \cite{SAS}, solvent-excluded (or Connolly)
 \cite{SES}, Gaussian-based \cite{ZAP}, spline-based \cite{Im1998}, and differential geometry surfaces \cite{Bates08, Bates09, Chen2010, Cheng07, Dzubiella06, Dzubiella06b}.
All of these interface models represent atoms as spheres and require information about the radii of these spheres.
These radii are generally assigned to sets of atoms based on their ``type'' as determined by the local molecular connectivity.
Unlike atomic charges, there are relatively few sets of atom types used to assign radii \cite{NichollsMGCBCP08, PARSE}.
These radii parameters are determined by optimization of properties such as solvation energy against experimental data \cite{NichollsMGCBCP08, PARSE}.
Additionally, many of these models also require information about solvent characteristics, generally in the form of a solvent radius, characteristic solvent length scales, or bulk solvent pressure/surface tension properties.

\revision{The intent of this study is to provide a way for developers of implicit solvent model parameter sets to understand the sensitivity of their target properties (solvation energy) on underlying choices for solute radius and charge parameters.}
In the present work, we \revision{present a new method to quantify the uncertainty in solvation energy calculated by the Poisson equation and induced by the uncertainty of the input radii and charge parameters.}
In particular, we construct two surrogate (or statistical regression) models of the solvation energy in terms of the radii and the atomic charges, respectively.
These surrogate models enable us to estimate the solvation energy with different input parameters quickly and to evaluate the statistical information of the target properties (e.g., probability density function) efficiently.
We model the input parameters as independent (i.i.d.) Gaussian random variables with different means and standard \revision{deviations; however, other probability distributions can also be used.}
To construct the surrogate of the Poisson model, we use a generalized polynomial chaos (gPC) \cite{GhanemS91, XiuK02} expansion to represent the dependence of the solvation energy on uncertain parameters such as the atomic charge and radii.
The efficacy of the gPC method for elliptic problems such as the Poisson
equation has been extensively studied with robust results for its efficiency and accuracy \cite{TodorS07, BabuskaNT10}.
This approach is straightforward to apply to the relatively low-dimensional parameter sets.
However, the main challenge of applying this method to implicit solvent calculation parameter uncertainty is the high-dimensionality of parameter sets
(especially the atomic charges):  the surrogate models require more basis functions and, therefore, more expansion coefficients need to be identified.
To address this challenge, we adopt a compressive sensing method combined with the rotation-based sparsity-enhancing method first proposed by Lei et al.~\cite{LeiYZLB15} and extended by Yang et al.~\cite {YangLBL16}, which enable us to construct the surrogate with relatively few sample outputs of the numerical Poisson solver.

\section{Methods}
\label{sec:method}

We demonstrated the framework using a test set of 17 compounds from the SAMPL computational challenge for solvation energy prediction \cite{NichollsMGCBCP08} (see Table \ref{tab:comp_list}).
\revision{This set was chosen to demonstrate the uncertainty quantification framework on several different molecules; however, it was not chosen to calculate statistics over this small set.}
\begin{table}
  \centering
  \caption{List of 17 compounds from the SAMPL computational challenge for solvation energy prediction {with solvation energy \cite{NichollsMGCBCP08} and solvent accessible volume from APBS \cite{BakerSJHM01}.}}
	\begin{tabular}{C{2em}L{16em}C{8em}C{8em}}
    \hline \hline
    Ind. & \qquad Compound & Solvation energy (kJ/mol) & Molecular volume ({\AA}$^3$) \\
    \hline
    1 & glycerol triacetate  & -36.99 & 215.80\\
    2 & benzyl bromide       & -9.96  & 126.98 \\
    3 & benzyl chloride      & -8.08  & 124.66 \\
    4 & $m$-bis(trifluoromethyl)benzene & 4.48 & 290.91  \\
    5 & $N,N$-dimethyl-$p$-methoxybenzamide  & -46.07 &  188.44 \\
    6 & $N,N-4$-trimethylbenzamide  & -40.84 & 179.62 \\
    7 & bis-$2$-chloroethyl ether   & -17.70 & 121.45\\
    8 & $1,1$-diacetoxyethane       & -20.79 & 148.50 \\
    9 & 1,1-diethoxyethane          & -13.72 & 132.75\\
    10 & $1,4$-dioxane              & -21.13 & 87.86\\
    11 & diethyl propanedioate       & -25.10 & 165.44\\
    12 & dimethoxymethane            & -12.26 & 81.90\\
    13 & ethylene glycol diacetate    & -26.53 & 148.10\\
    14 & $1,2$-diethoxyethane     & -14.81 & 132.91\\
    15 & diethyl sulfide  & -6.49 & 108.15\\
    16 & phenyl formate    & -15.98 & 126.25 \\
    17 & imidazole  & -41.05 & 67.36\\
  \hline \hline
  \end{tabular}
  \label{tab:comp_list}
\end{table}
\revision{We use this subset of the SAMPL data to demonstrate the use of our method to quantify uncertainty in solvation energy due to implicit solvent parameter uncertainty.}

\subsection{Uncertain parameters}

Many parameterization approaches for atomic charge use ESP (electrostatic potential) \cite{esp} or related methods (e.g., RESP \cite{resp}).
These methods optimize atomic charges by least-squares fitting of the charges' Coulombic potential to the electrostatic potential obtained from quantum mechanical calculations.
This under-determined optimization is performed subject to various constraints, including the requirement that the atomic charges sum to the integer formal charge of the molecule.
More specifically, the calculated ESP $\hat V_i$ at the $i$-th grid point is the electrostatic potential given by Coulomb's law summed over the charge $q_j$ at the centers of the $j$-th atoms.
Least-squares fitting is performed by minimizing $\sum_i (V_i-\hat V_i)^2$ with constraints, where $V_i$ is the electrostatic potential computed by \emph{ab initio} calculations.
Least-squares fitting implies a Gaussian noise model wherein the atomic charges $q_j$ can be modeled as Gaussian random variables \revision{as done in this study.}

In the present work, we modeled the uncertainty in atomic charges by considering atomic charges obtained by 11 different approaches: AM1BCC~\cite{am1bcc}, CHELP~\cite{chelp}, CHELPG~\cite{chelpg}, CM2~\cite{cm2}, ESPMK~\cite{esp}, Gasteiger~\cite{gasteiger}, PCMESP~\cite{pcmesp}, Qeq~\cite{qeq}, RESP~\cite{resp}, MMFF94~\cite{mmff94}, Mulliken~\cite{mulliken}.
The Hartree-Fock method and the 6-31G*basis set were used to optimize molecular geometries.
\revision{The methods we selected here are popular for implicit solvation models and all-atom approaches.
Although many of these charge methods are used in all-atom simulations, implicit solvent models have been used with several of them, including RESP and ESP(MK) \cite{Knight2011Surveying, PARSE}, AM1-BCC \cite{Knight2011Surveying}, Mulliken \cite{Hou2010Implicit}, CHELPG \cite{Ginovska2008ChargeDependent, Hou2010Implicit}, Gasteiger \cite{Czodrowski2006Development}, Qeq \cite{Yang2006Atomic}, etc.}

We have \emph{assumed} that the variation of atomic charges across different methods can be modeled by a Gaussian random field with covariance kernel
\begin{equation} \label{eq:cov}
	\text{Cov}(\bm x_i,\bm x_j)=\eta_i\eta_j\exp\left(-\dfrac{\Vert\bm x_i-\bm x_j\Vert_2^p}{\theta}\right),
\end{equation}
where $\eta_i$ is the standard deviation of the $i$-th atomic charge, $\bm x_i$ is the position of the $i$-th atom, and $0<p<2$.
\revision{The least-squares nature of most charge fitting methods makes Gaussian variables a natural choice; however, other probability distributions can also be used.}
We used atomic charges from 11 different methods to estimate $\eta_i$ and then used the maximum likelihood estimate (MLE) method to estimate $\theta$ and $p$.
Since the sum of $N_A$ charges in a molecule is constrained (to its formal molecular charge $Q \in \mathbb{Z}$), we modeled the Gaussian random field with $N_A-1$ atoms by removing the last hydrogen in the PDB file.
Additionally, we use symmetry in the molecular structure to reduce the number of independent atomic charge types before applying the MLE to identify the random field.
For example, in a benzene, there is only one type of carbon and one type of hydrogen due to the symmetry of this molecule.
Therefore, we considered the charges of its atoms as a Gaussian random field with only two entries instead of 12 ones (the total number of atoms in benzene).

After obtaining the covariance matrix by integrating across methods, we represented the atomic charge as
\begin{equation}\label{eq:charges}
	\bm q = \mean{\bm q} + \tensor L_c\bm\gamma,
\end{equation}
where $\bm q=(q_1,q_2,\cdots,q_{_{N_A-1}})$ are the atomic charges, $\mean{\bm q}$ is the mean of $\bm q$ estimated from the 11 different charge values, $\bm\gamma=(\gamma_1,\gamma_2,\cdots,\gamma_{_{N_A-1}})$ are i.i.d.\ zero-mean unit-variance Gaussian random variables, and $\tensor L_c$ is a lower triangular matrix from the Cholesky decomposition of the covariance matrix (Eq.~\ref{eq:cov}).
We note that for the atoms in the test set used in the present work, the covariance matrices of these random field are almost diagonal: the off-diagonal entries are smaller than $10^{-12}$.
This suggests the correlation between atomic charges is \revision{effectively} removed during their symmetry-based grouping.
The atomic charge for the remaining atom is obtained by summation of the other random charge variables based on the constraint $q_i = Q - \sum_{j \neq i}^{N_A} q_j$.

Similarly, we used multiple force fields (ZAP-9 \cite{NichollsMGCBCP08}, OPLSAA \cite{Jorgensen_JACS_1996}, Bondi \cite{BONDI} and PARSE \cite{PARSE}) to model uncertainty in the radii parameters in the same manner.
Although radii are non-negative, we did not explicitly impose constraints on the radii.
After obtaining the covariance matrix, we represented the radii as
\begin{equation}\label{eq:radii}
	\bm \sigma = \mean{\bm \sigma} + \tensor L_r\bm\zeta,
\end{equation}
where $\bm\sigma=(\sigma_1,\cdots,\sigma_{N_A})$, $\sigma_i$ is the radius of atom (type) $i$, $\bm\zeta=(\zeta_1,\cdots, \zeta_{N_A})$ are independent zero-mean unit-variance Gaussian random variable and $\tensor L_r$ is a lower triangular matrix from the Cholesky decomposition of the covariance matrix.
\revision{The small number of radii sets makes the selection of a probability distribution somewhat arbitrary.
We have \emph{assumed} that the radii follow a Gaussian distribution; however, other probability distributions can also be used.}
We note that the standard deviations here are smaller than $10\%$ of the mean values which implies very low probabilities for unphysical negative radii values.
Therefore, by employing truncated Gaussian random variables within 4 standard deviations (capturing more than 99.99\% of the probability), we guaranteed that the radii are always positive and that the distributions of the truncated Gaussian variables were almost identical to the original Gaussian variates.
\revision{We note that with this setting, no model assigns zero radius to protons or other atoms.}

Although we use $\gamma$ and $\zeta$ to denote the random variables used for modeling the uncertainties in $q_j$ and $\sigma_j$, in what follows, we still use $\bx=(\xi_1,\xi_2,\cdots)$ to denote general uncertain inputs when introducing the algorithm and reporting results.

\subsection{Solvation energy surrogate models}

We used generalized polynomial chaos (gPC) expansions as surrogate models for the solvation energy.
The goal of surrogate construction is to estimate the variations in quantities of interest, such as solvation energy, much more efficiently than solving the original problem, such as solving the Poisson equation.
The details for these expansions are provided in Supporting \revision{Material}.

\subsection{Poisson equation solver}
We used the Adaptive Poisson-Boltzmann Solver (APBS) \cite{BakerSJHM01} to solve the Poisson equation for solvation energies.
\revision{Poisson calculations were performed with the finite difference solver using $97^3$ grids focused from a 25 \AA\ to a 13 \AA\ cubic domain.
Charges were discretized onto the grids using linear interpolation.
Boundary conditions were assigned using a sum of Coloumb potentials.
The molecular interior and solvent were assigned dielectric values of 2.0 and 78.0, respectively.
The solute-solvent boundary was defined using a ``Connolly'' molecular surface \cite{Connolly1985}.
Energies were calculated using the standard approach for Poisson-Boltzmann calculations \cite{Micu1997Numerical, Sharp1990Calculating}.}

\section{Results and discussion}

For each test case, we used Monte Carlo simulations to \revision{sample the parameter probability distributions and} generate $10,000$ samples of the input parameters $\bm\xi^q$ and then solved PB equation using APBS to obtain output samples of the solvation energy $E^q=E(\bm\xi^q)$.
\revision{We used these outputs as ground-truth reference solutions to examine the performance of the surrogate models; these outputs will be referred to as ``reference'' in the remainder of this paper.}
More precisely, given a surrogate model $\tilde E$, we use two different root-mean-squared error (RMSE) measures to examine its accuracy:
\begin{equation}
	RMSE_1 = \sqrt{\dfrac{\sum_{q=1}^{10000} \left(\tilde E(\bm\xi^q)-E^q\right)^2} {\sum_{q=1}^{10000} (E^q)^2}},\quad
	RMSE_2 = \sqrt{\dfrac{\sum_{q=1}^{10000} \left(\tilde E(\bm\xi^q)-E^q\right)^2} {10000}}.
\end{equation}
We also use box-whisker plots to demonstrate the statistics.
The line in the middle is the median of 16 molecules, the tops and bottoms of the boxes are 25th and 75th percentiles, and the whisker plots cover more than $99\%$ probability.

\subsection{Influence of radii uncertainties on solvation energies}
\label{subsec:radii}

\begin{figure}
	\centering
	\subfigure[]{\includegraphics[width=3in]{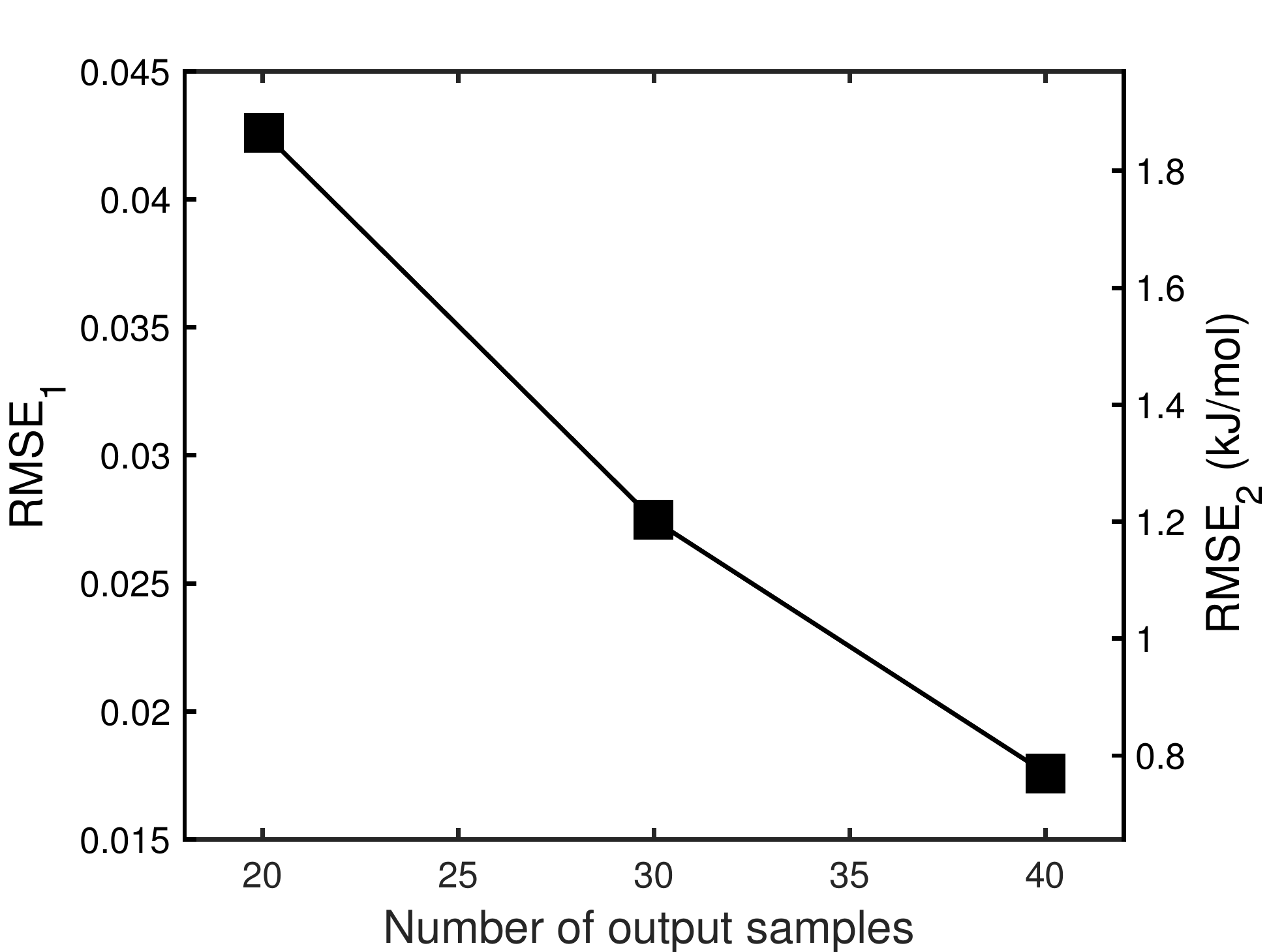}}
	\subfigure[]{\includegraphics[width=3in]{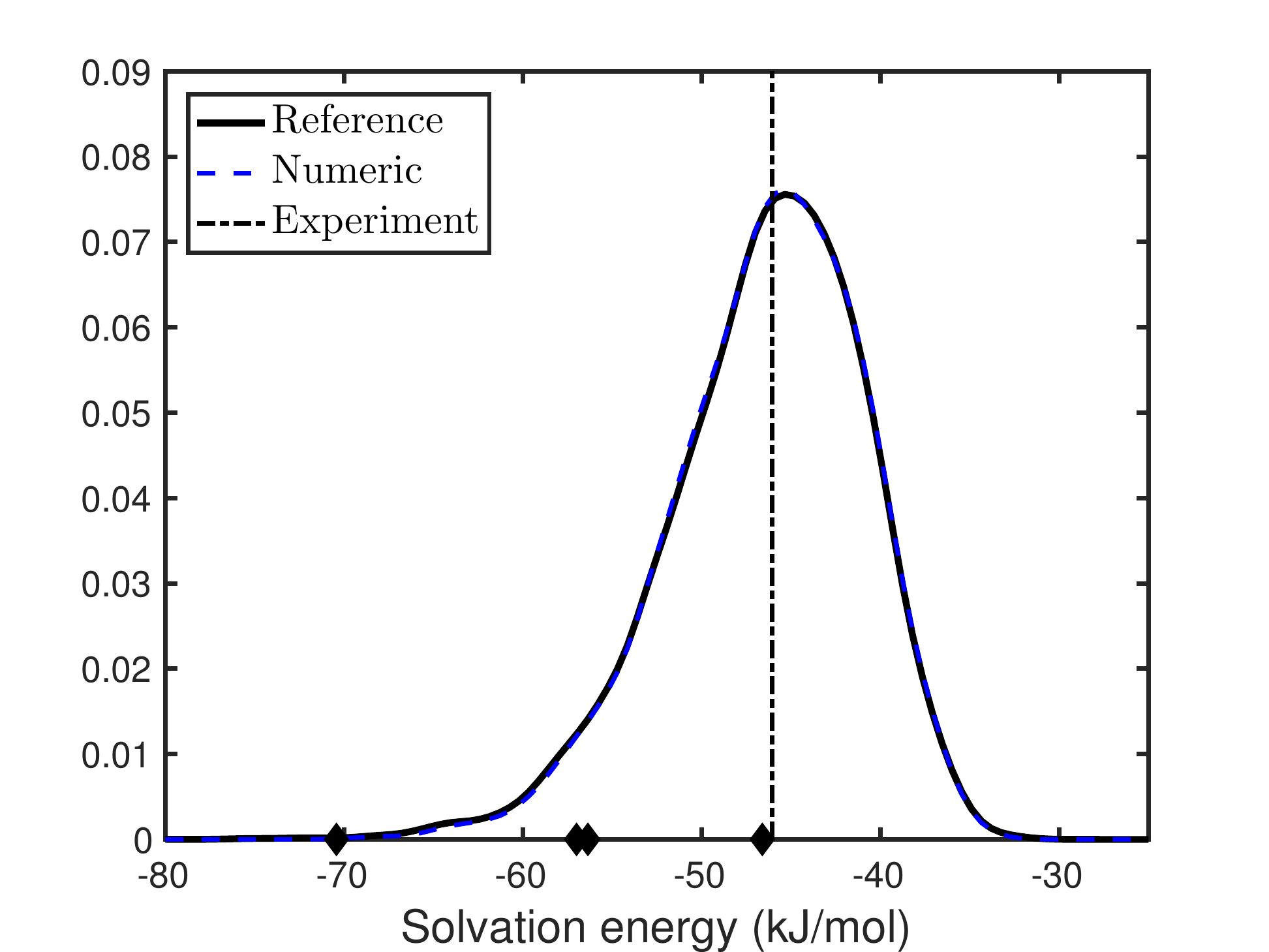}}
	\caption{Performance of the surrogate model for radii uncertainties for $N,N$-dimethyl-$p$-methoxybenzamide.
	(a): RMSE with different numbers of samples $M$.
	(b): comparison of the solvation energy PDFs estimated by the numerical surrogate method (``Numeric") based on $40$ output samples of APBS; dash line (``Experiment") is the experimental result; diamonds are the results by using radii from ZAP-9, Bondi, OPLSAA and PARSE, respectively.
	The diamond closest to the experiment was obtained from ZAP-9.}
	\label{fig:nnd_radii}
\end{figure}

We investigated the effect of the uncertainties in the radii with fixed atomic charges obtained from AM1-BCC \cite{am1bcc}.
As an example, there are eight different sets of radii for $N,N$-dimethyl-$p$-methoxybenzamide across the ZAP-9, Bondi, OPLSAA, and PARSE parameter sets, as shown in the support material.
We modeled the solvation as a function of eight i.i.d.\ Gaussian random variables.
We constructed gPC surrogate models with multi-variate normalized Hermite polynomials up to third order.
The surrogate model consisted of $C_{8+4}^4=495$ basis functions.
Figure~\ref{fig:nnd_radii}~(a) presents the RMSE obtained by our method with respect to different numbers of samples $E^q$.
Figure~\ref{fig:nnd_radii}~(b) compares the solvation energy probability distribution function (PDF) obtained by our method and the reference solutions.
The numerical results are obtained by constructing the surrogate model with the $36$ output samples first, then sampling the surrogate model $10,000$ times with random samples to estimate the PDF.
The reference solution is computed from the $10,000$ outputs of $E^q$.

\begin{figure}
\centering
	\includegraphics[width=3.6in]{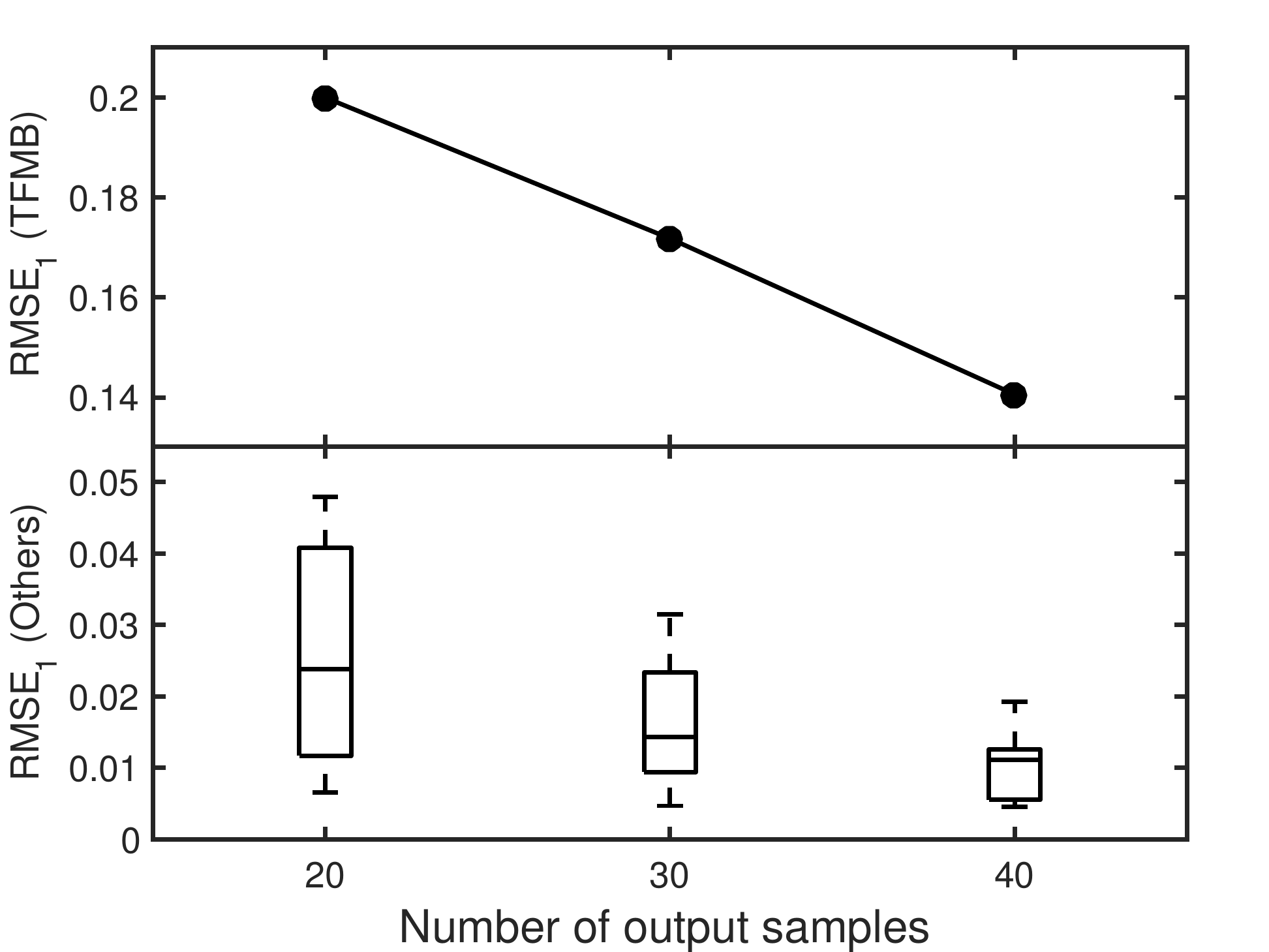}
	\caption{Performance of surrogate models with respect to number of samples.
	Circles are the $RMSE_1$ of $m$-bis-trifluoromethylbenzene (TFMB), box-whisker plots are the $RMSE_1$ of the remaining $16$ molecules. }
	\label{fig:radii_all}
\end{figure}

\begin{figure}
\centering
	\includegraphics[width=3.6in]{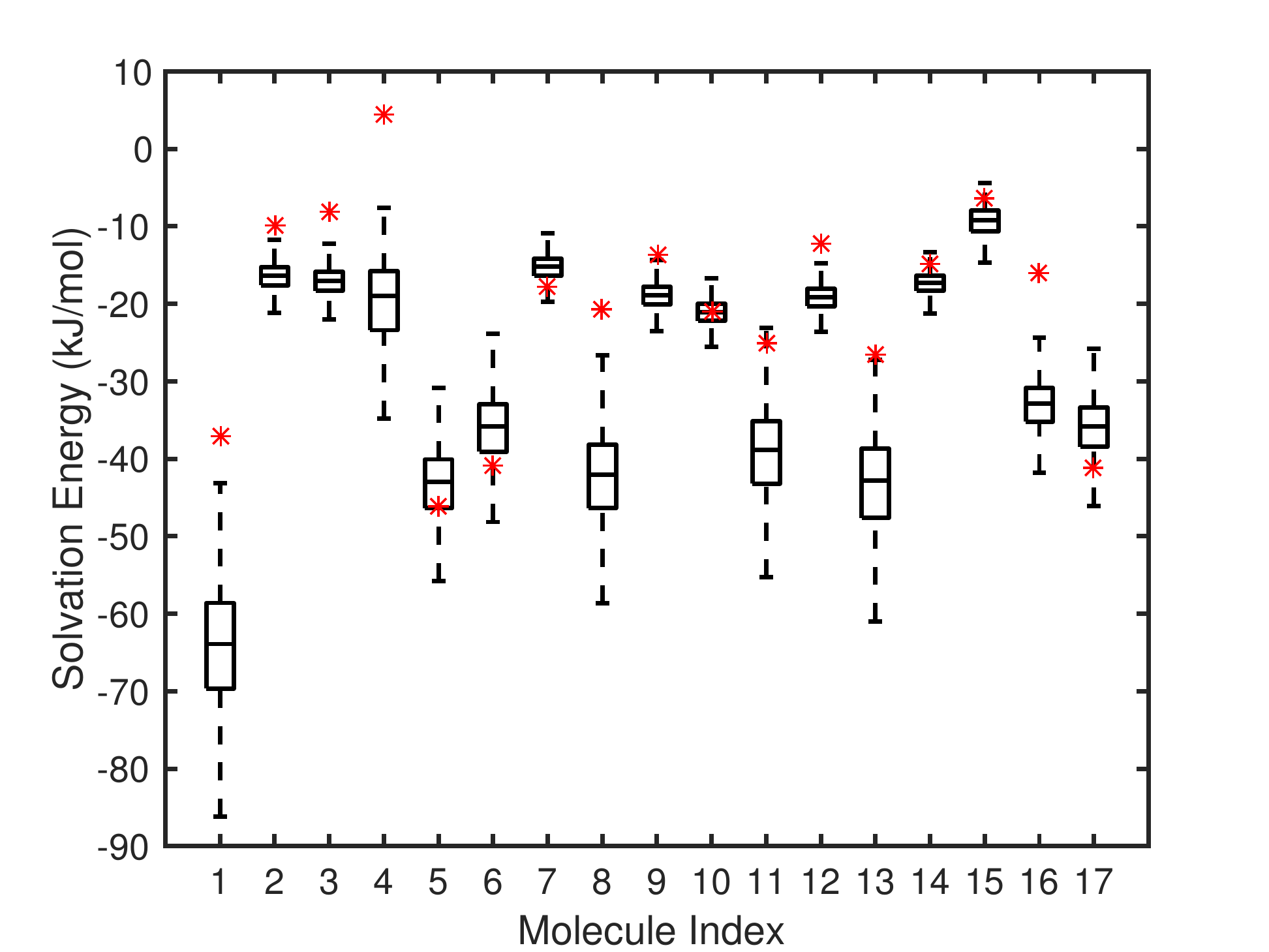}
	\caption{Influence of radii uncertainties on molecular solvation energies for the 17-molecule test set.
	The red stars are the experiment results.}
	\label{fig:radii_box_all}
\end{figure}

We performed the same analysis for all the molecules in the test set and present the results in Figure~\ref{fig:radii_all}.
For most molecules, we can build an accurate surrogate model (RMSE$<0.05$) for the solvation energy with only a few samples (less than $40$) of the input parameters.
However, $m$-bis-trifluoromethylbenzene (TFMB) required significantly more samples.
In particular, the RMSE for the TFMB solvation energy surrogate model was close to $0.15$ with $40$ samples and required $100$ samples to reduce the RMSE to less than $5\%$.
This variability arises from the radius of fluorine: in the ZAP force field it is 2.4 \AA; however, it is only $\sim 1.4$ \AA\ for the other force fields.
Hence, the standard deviation of this radius is around $25\%$ of the mean and fluorine requires more terms in the surrogate model for an accurate description and therefore more samples to parameterize those terms.
The influences of the uncertainties in the input radii on the solvation energy for each molecule are demonstrated in box-whisker plots in Figure \ref{fig:radii_box_all}.
The experiment results are presented for comparison.
We note that some experiments results are ``outliers" of the box-whisker plots, this is because that the atomic charges are computed from AM1BCC for the purpose of fixing the atomic charges and it does not guarantee that the computed solvation energy is sufficiently close to the experiment results.
For example, for the $m$-bis(trifluoromethyl)benzene AM1BCC charges yield negative solvation energy while the experiment result is positive.

\subsection{Influence of atomic charge uncertainties on solvation energies}

\begin{figure}
	\centering
	\subfigure[]{
	\includegraphics[width=3in]{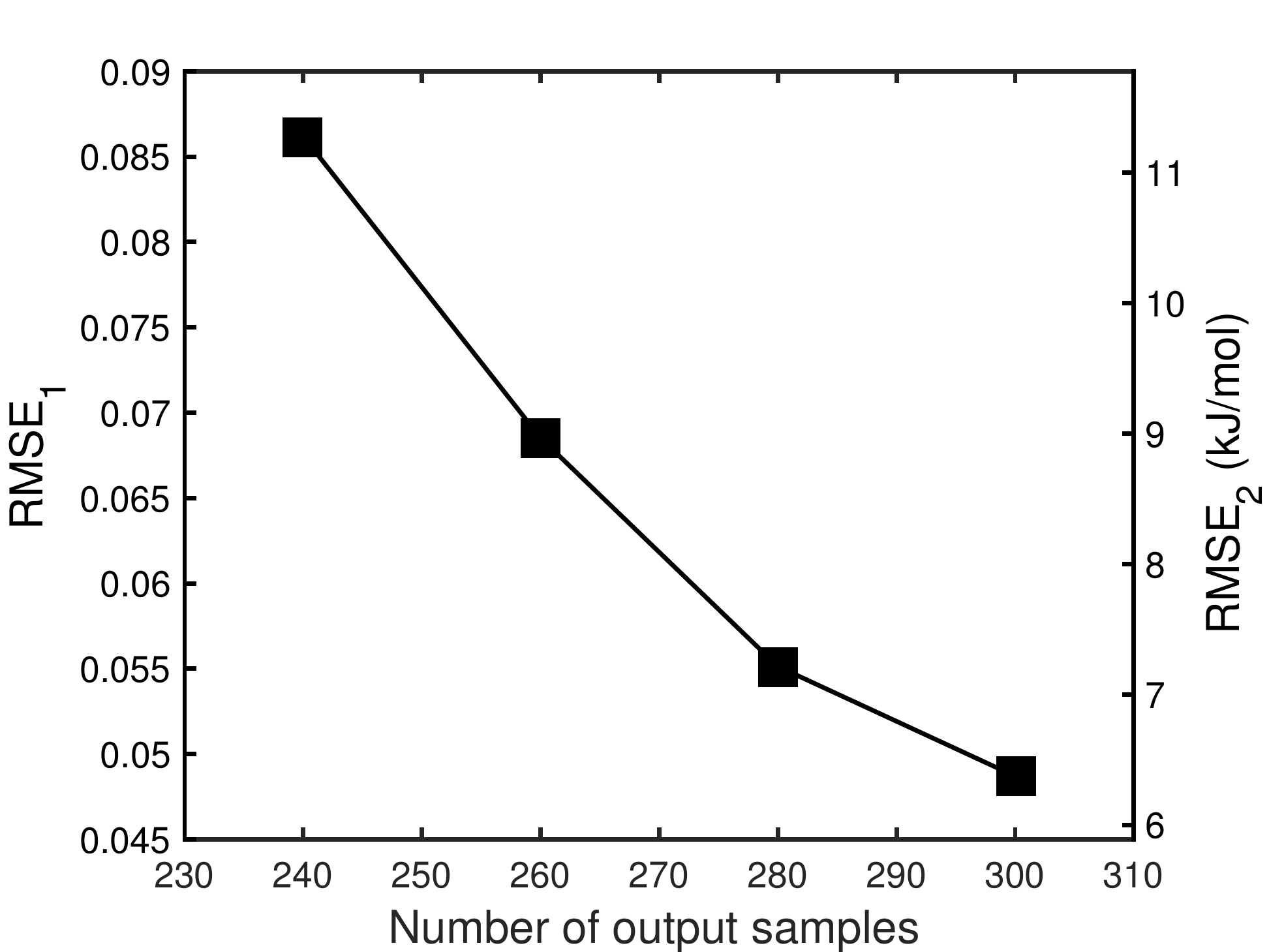}}
	\subfigure[]{\includegraphics[width=3in]{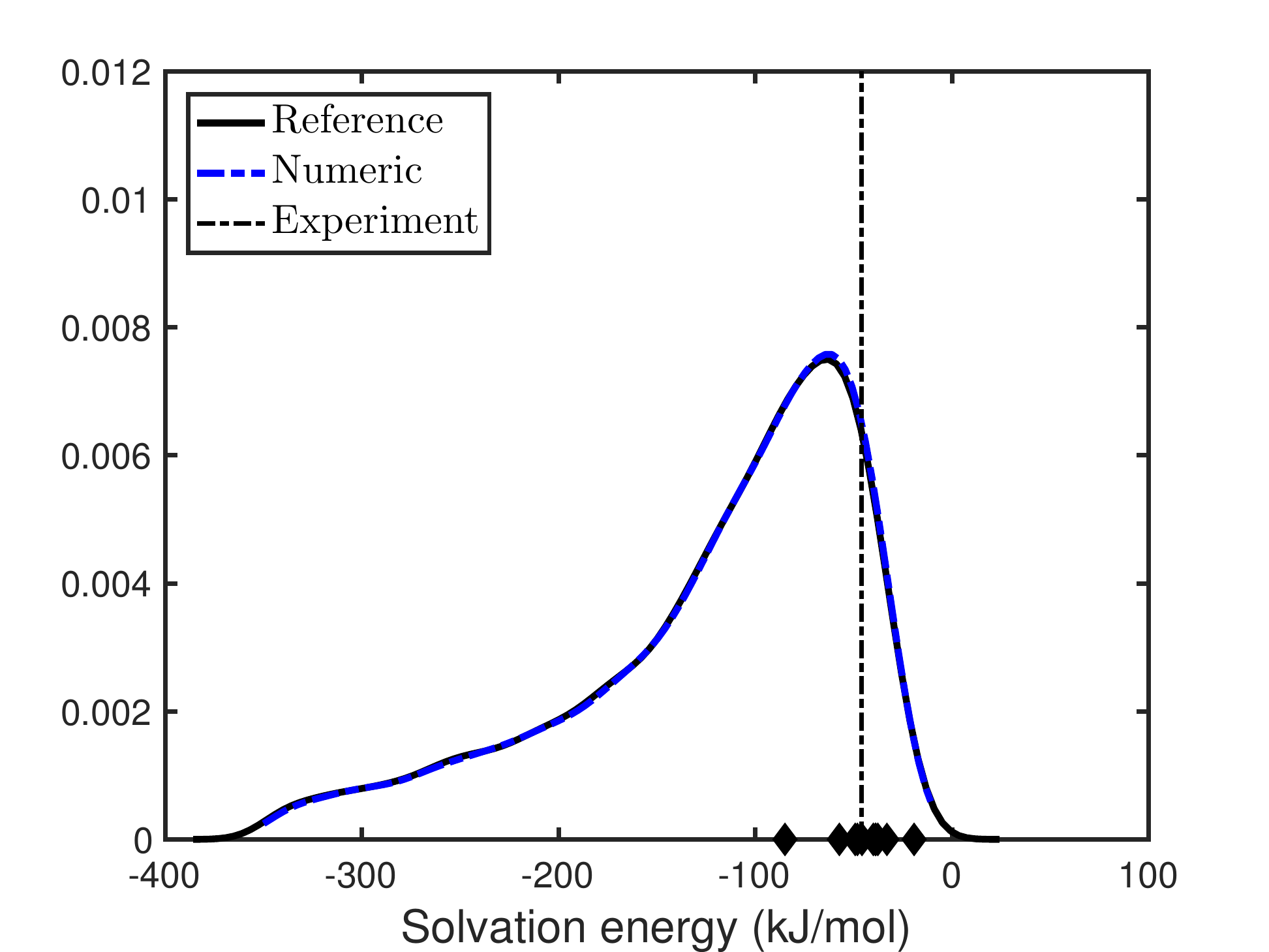}}
  \caption{Performance of surrogate models for charge uncertainties for $N,N$-dimethyl-$p$-methoxybenzamide.
	(a): RMSE for surrogate model with different number of output samples.
    (b): comparison of the PDFs estimated by the numerical surrogate method (``Numeric") based on $300$ output samples of APBS; dash line (``Experiment") is the result by the experiment; diamonds are results by using atomic charges from AM1BCC, CHELP, CHELPg, CM2, ESPMK, Gasteiger, PCMESP, Qeq, RESP, MMFF94, Mulliken.}
	\label{fig:nnd_charge}
\end{figure}

We also examined the influence of charge perturbation for solvation energy calculations with fixed radii (ZAP-9).
As an example, there are 14 different types of atoms in $N,N$-dimethyl-$p$-methoxybenzamide as shown in Supporting Material.
We note that we model the surrogate with 13 inputs due to the constraint on the summation of the charges.
The mean and standard deviation are computed from the results of 11 different charge fitting approaches.
We used no more than 3000 multi-variate normalized Hermite polynomials (up to fourth order) in the gPC surrogate model for $E_g$ for all the molecules.
We use $N,N$-dimethyl-$p$-methoxybenzamide as an example.
Figure~\ref{fig:nnd_charge}~(a) presents the RMSE obtained by our method with respect to different numbers of samples $E^q$.
It illustrates that 300 output samples are needed to reduce the RMSE to less than $5\%$.
Figure~\ref{fig:nnd_charge}~(b) compares the PDF obtained by our method and the reference solution.
The numerical results are obtained by constructing the surrogate model with the $300$ output samples first, then sampling the surrogate model $10,000$ times with random samples to estimate the PDF.
The reference solution is computed from the $10,000$ outputs of $E^q$.

\begin{figure}
\centering
	\includegraphics[width=3.6in]{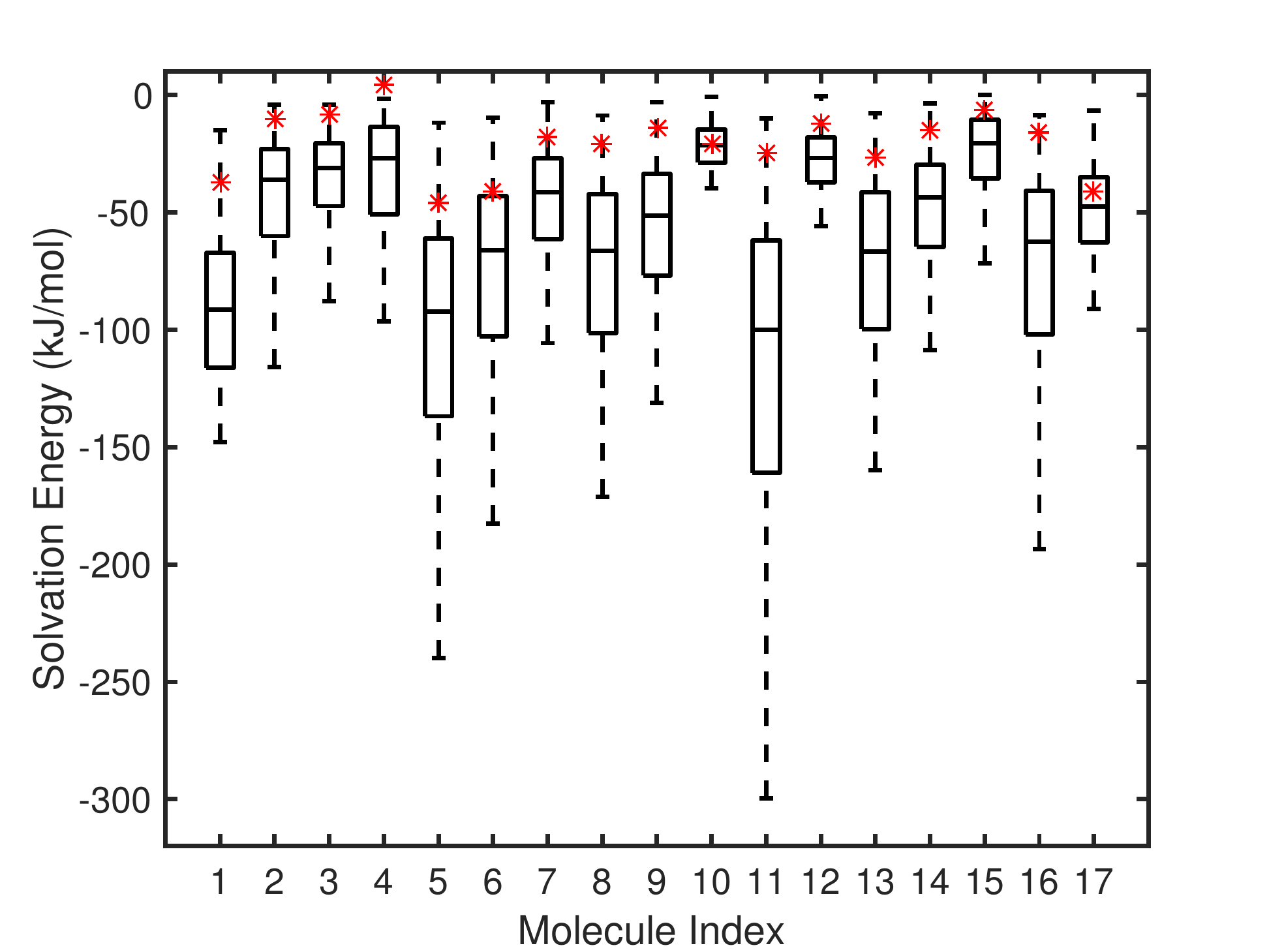}
	\caption{Results of atomic charge uncertainties.
	Box-whisker plots demonstrating the uncertainties in the numerical results of the solvation energy for 17 compounds.
	The red stars are the experiment results.}
	\label{fig:charge_box_all}
\end{figure}

\begin{figure}
	\centering
	\includegraphics[width=4in]{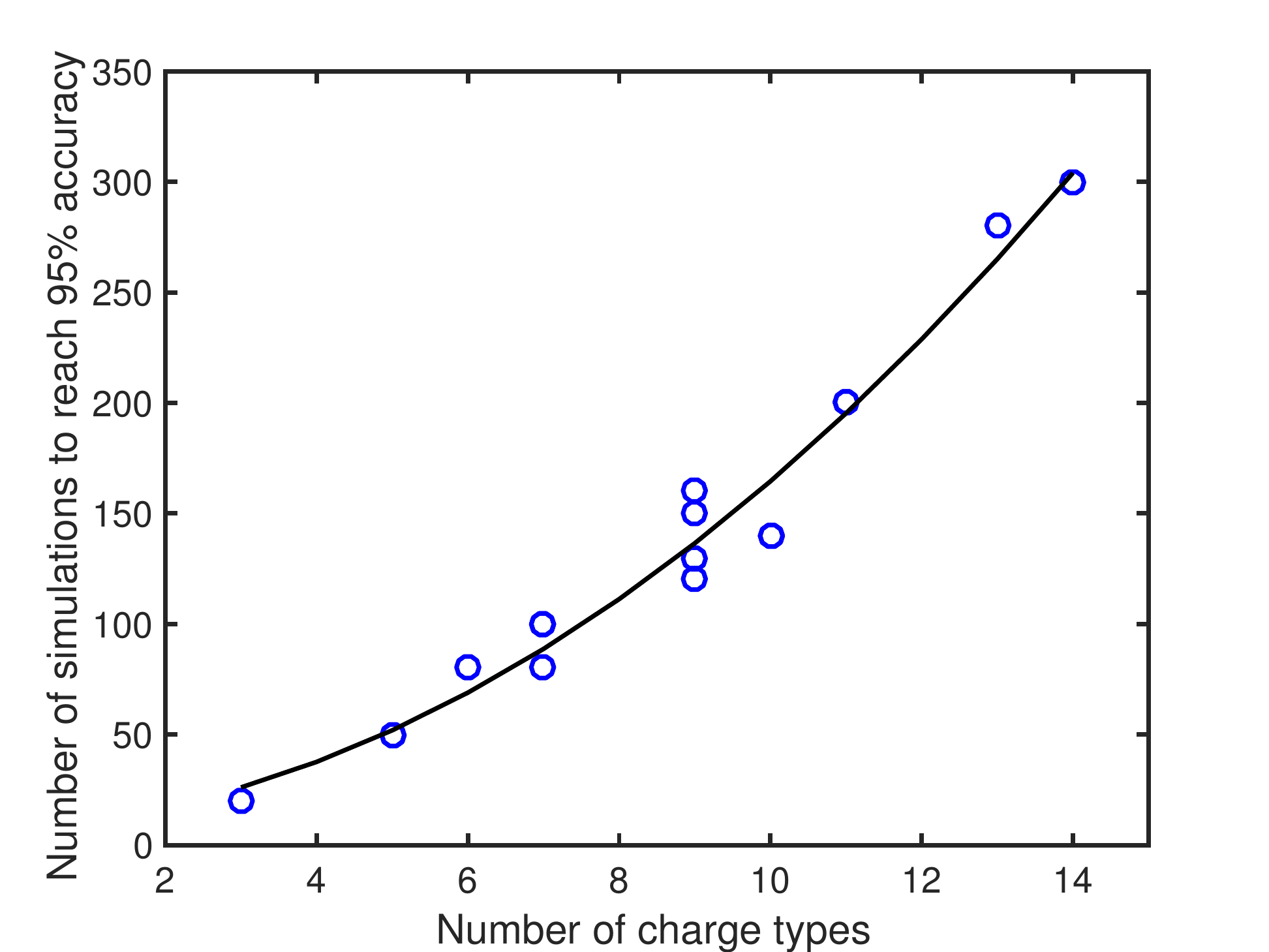}
	\caption{ ``$\circ$" : number of output samples \revision{needed} to construct a surrogate model with RMSE less than $5\%$ with respect to the number of atom types; ``-" is the best-fit curve $1.4 x^2 + 1.9 x + 7.9$.}
	\label{fig:charge_num_sam}
\end{figure}

The influences of the uncertainties in the input atomic charges on the solvation energy for each molecule are demonstrated in Figure \ref{fig:charge_box_all}.
For most molecules, the experiment results lie in the whisker plots and some of them are in the box.
We also present the number of output samples needed to construct a surrogate with RMSE less than $5\%$ with respect
to the number of atom types in Figure \ref{fig:charge_num_sam}.

\subsection{Combined influence of radius and atomic charge uncertainties}
\revision{ We chose the charge methods and radii based on their popularity in the implicit solvent community.
Not all of the radii and charges examined in this study would be expected to give accurate answers when used together.
We could have chosen a more constrained set; however, we chose this diverse set as a more challenging example to test our method to illustrate our approach across significant variation in parameter values.}
Comparing the PDFs in Figures~\ref{fig:nnd_radii}~(b) and \ref{fig:nnd_charge}~(b), we notice that the uncertainty in the solvation energy induced by the atomic charges is stronger than that induced by the radii.
\revision{A similar observation has been made previously by Chakavorty et al.~\cite{Chakavorty2016} who also noted that conformation introduces another important source of uncertainty across force fields.
We have investigated the influence of conformational uncertainty on solvation in a previous paper using similar methods~\cite{LeiYZLB15}; however, combining parameter and conformational uncertainty is outside the scope of this manuscript.}

\begin{figure}
	\centering
	\subfigure[]{
	\includegraphics[width=3in]{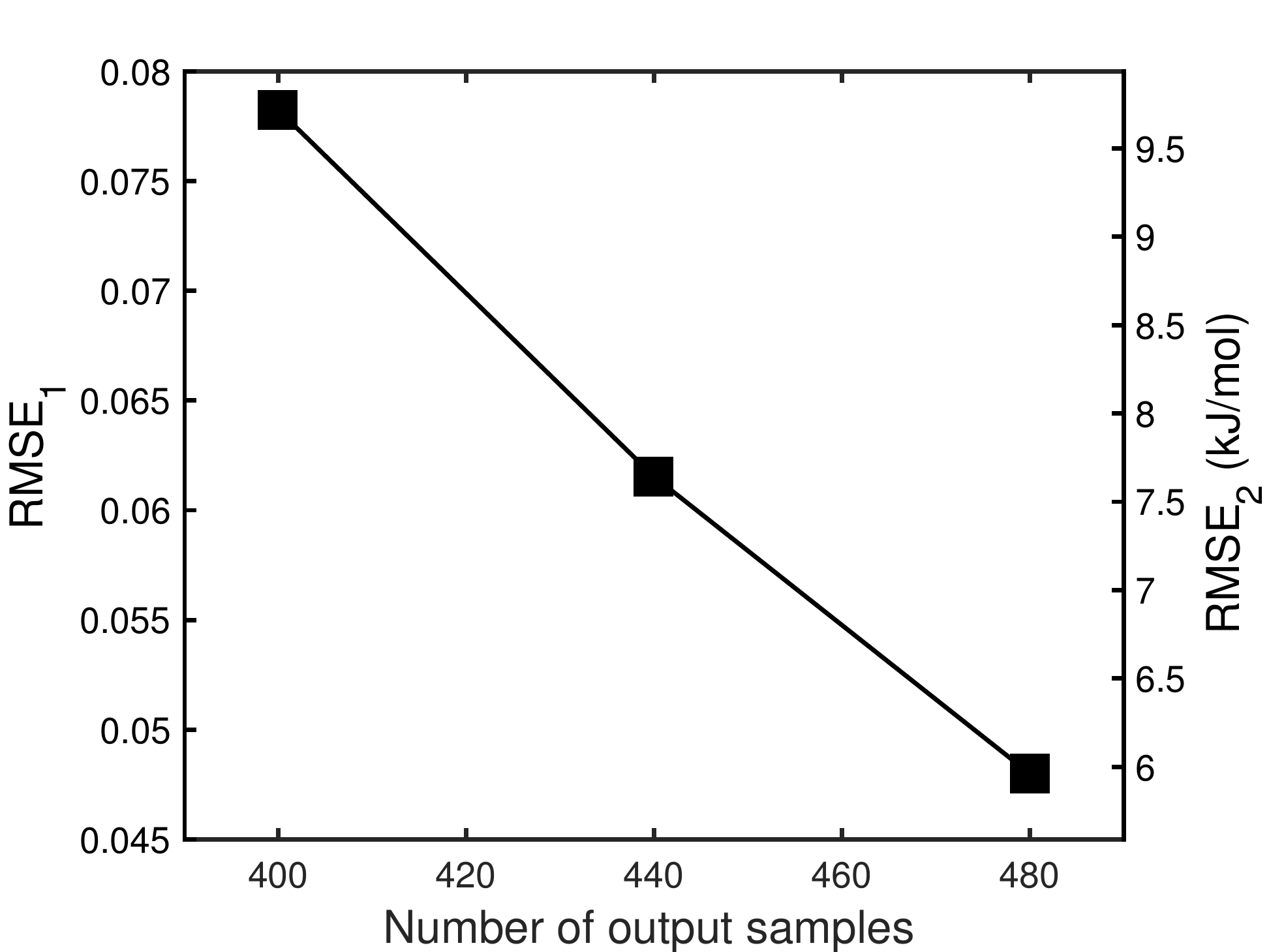}}
	\subfigure[]{\includegraphics[width=3in]{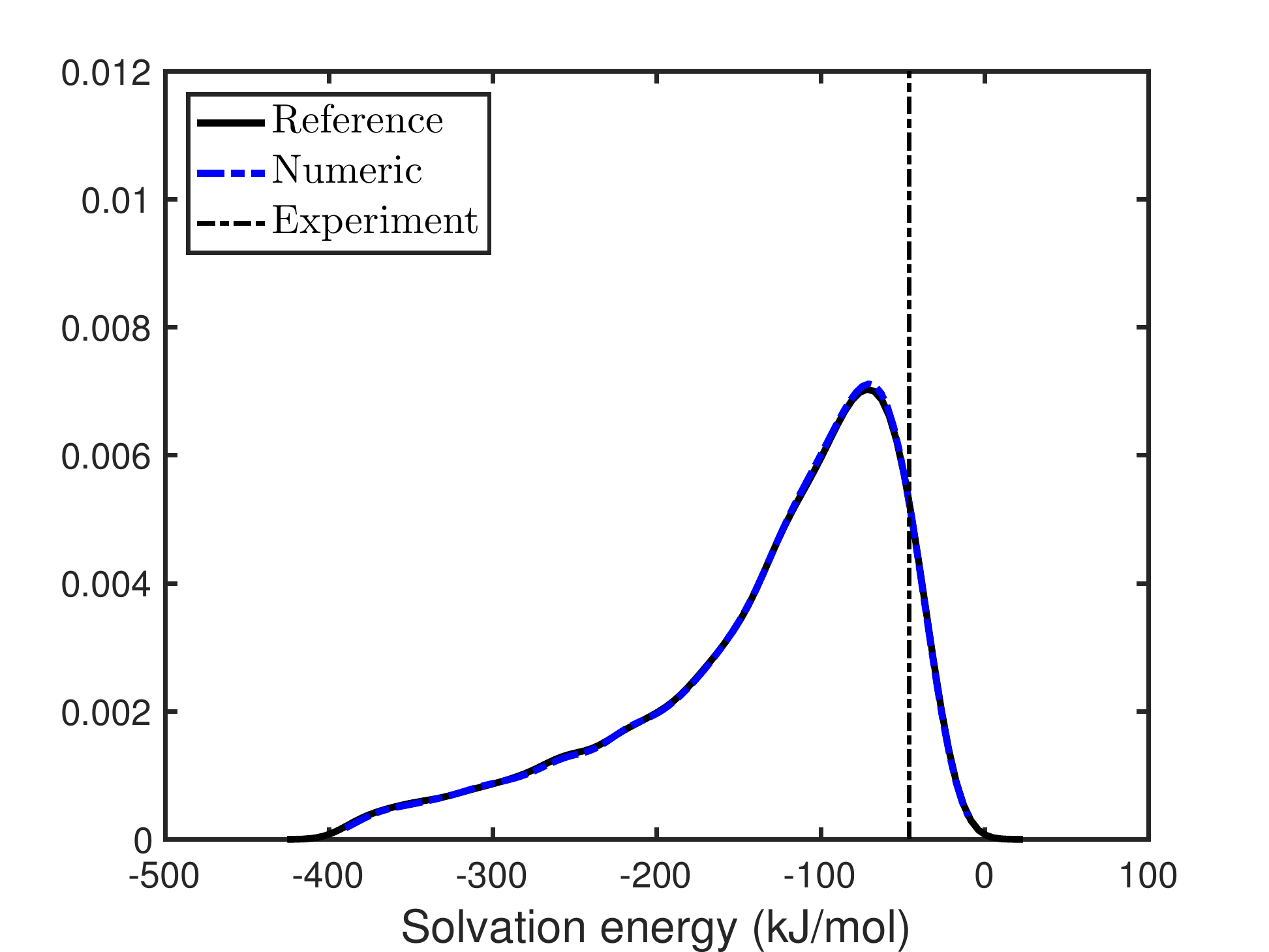}}
	\caption{Results of radii and charges uncertainties for $N,N$-dimethyl-$p$-methoxybenzamide.
	(a): RMSE with different number of output samples $M$.
	(b): comparison of the PDFs estimated by the numerical method (``Numeric") based on $480$ output samples of APBS; dashed line (``Experiment") is the experimental result.}
	\label{fig:nnd_charge_radii}
\end{figure}

The atomic charges vary significantly across different methods while the variation in the radii is much smaller.
To understand the combined influence of charges and radii on solvation energies, we modeled the correlated uncertainties for these two types of parameters can be modeled with i.i.d.\ Gaussian random variables.
We use $N,N$-dimethyl-$p$-methoxybenzamide as an example.
480 output samples are needed to reduce the RMSE to less than $5\%$.
Figure~\ref{fig:nnd_charge_radii}~(a) presents the RMSE obtained by our method with respect to different numbers of samples $E^q$.
Figure~\ref{fig:nnd_charge_radii}~(b) compares the PDF obtained by our method and the reference solution.
The numerical results are obtained by constructing the surrogate model from the $480$ output samples and then sampling the surrogate model $10,000$ times with random samples to estimate the PDF.
The reference solution is computed from the $10,000$ outputs of $E^q$.
Not surprisingly, the number of output samples needed to construct an accurate surrogate increases as we take into account both uncertainties in the charges and radii.
The shape of the solvation energy changes PDF also slightly as the radii variation of the radii across different methods are much smaller than charge variations.

\begin{figure}
\centering
	\includegraphics[width=3.6in]{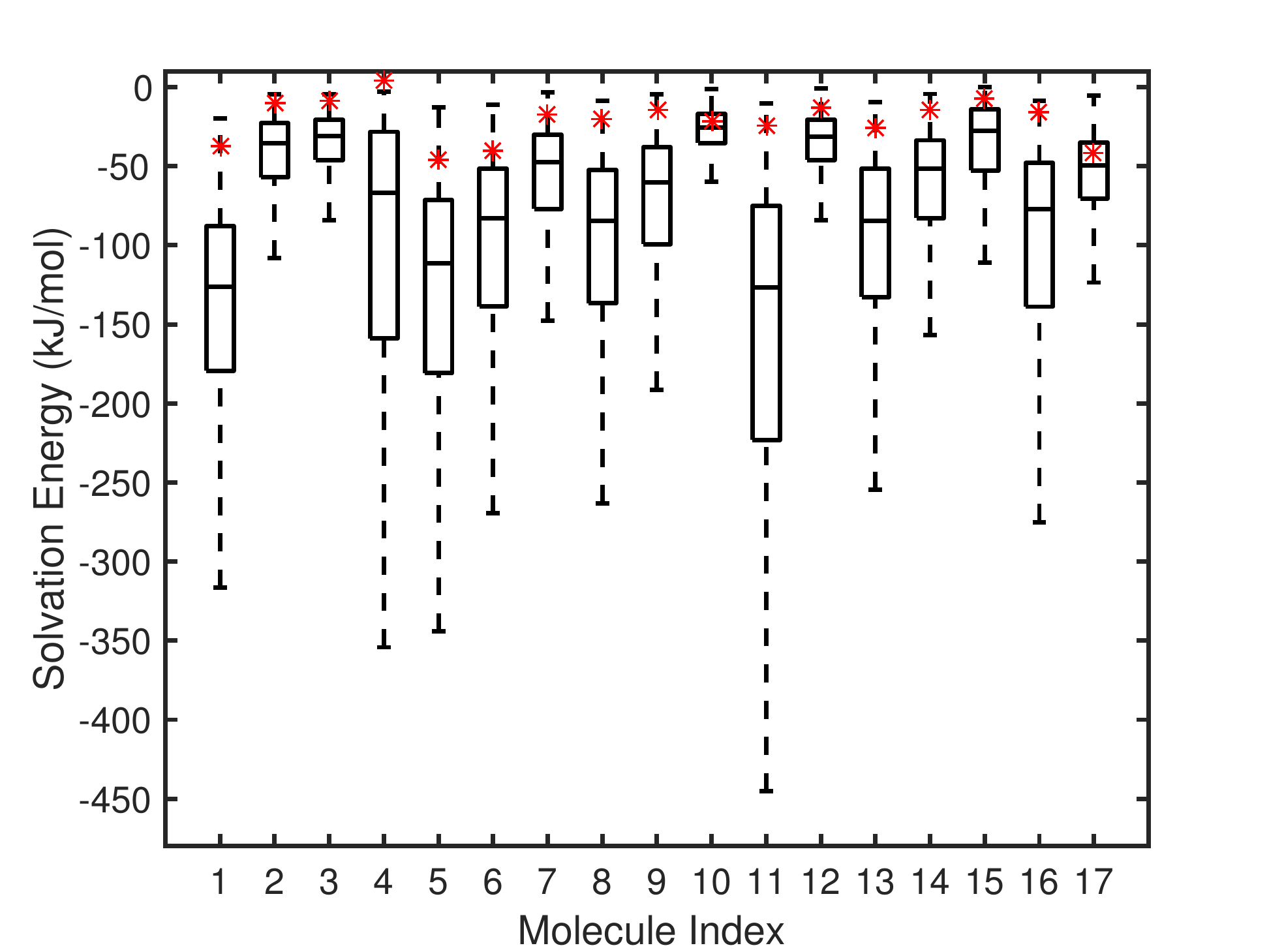}
	\caption{Results of radius and atomic charge uncertainties.
	Box-whisker plots demonstrating the uncertainties in the numerical results of the solvation energy for 17 compounds.
	Red stars are the experiment results.}
	\label{fig:rc_box_all}
\end{figure}

\begin{figure}
	\centering
	\includegraphics[width=4in]{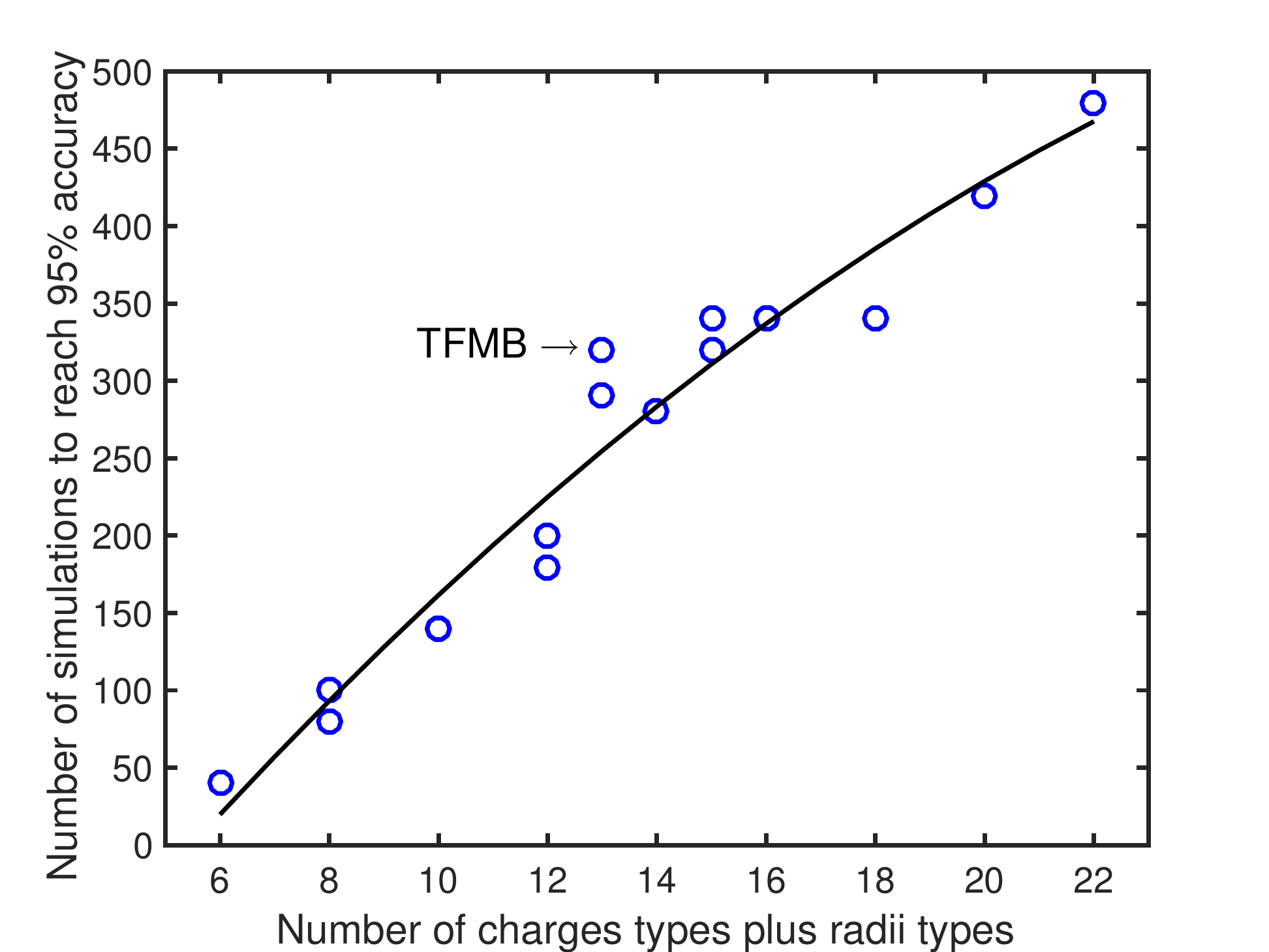}
	\caption{ ``$\circ$" : number of output samples \revision{needed} to construct a surrogate model with RMSE less than $5\%$ with respect to the number of atom charge types plus radius types; ``-" fitting curve $-0.6 x^2 + 45 x - 188$.}
	\label{fig:charge_radii_num_sam}
\end{figure}

The influences of the uncertainties in the input atomic charges on the solvation energy for each molecule are \revision{demonstrated} in Figure~\ref{fig:rc_box_all}.
This figure is similar to Figure~\ref{fig:charge_box_all} since the uncertainties in the atomic charges dominate the results.
Figure \ref{fig:charge_radii_num_sam} shows the number of output samples needed to construct a surrogate with less than $5\%$ RMSE for all 17 molecules in the test set.
This figure illustrates the approximately quadratic scaling with the respect to the number of atom types in the molecule.

\section{Conclusions}

\revision{We have developed a new method for quantifying the uncertainty associated with parameterization of implicit solvent models.
In particular, we} used a newly developed extension of compressive sensing method to construct surrogate models of solvation energy based on gPC expansions.
These surrogate models allow us to efficiently and accurately estimate the variation in solvation energy due to uncertainty in \revision{charge and radius parameters}.
\revision{In this initial work, we used statistical distributions for radius and charge variation based on the observed differences in the parameter sets.
However, in future studies, it may be useful to use the uncertainty quantification approach presented here with more physically motivated models that address the underlying uncertainties in determining charge and radius parameters.}
Our results demonstrate that for the data sets used in the present work, the variation of radii across different approaches are small.
\revision{On the other hand, the variations of the atomic charges obtained by different methods are much larger, so that the number of output samples needed for accurate UQ analysis requires are much larger, growing quadratically with respect to the number of atom types.}
This framework can be applied to estimate the statistics (e.g., mean, variance), PDF, confidence interval, Chernoff-like bounds \cite{rasheed2016}, etc. of solvation computing and other chemical computing when the inputs are uncertain.
The current study focused on uncertainty in solute charges and radii; however, this framework could also be applied to other solvation model characteristics such as dielectric coefficient, solvent radius, and biomolecular surface definition.
\revision{Likewise, this approach could also be used for quantities of interest other than solvation energy; e.g., dipole moments, titration states, etc.}

In the future, we anticipate that this approach could be used for a much wider range of force field parameterization activities, including both coarse-grained and atomistic representations of biomolecules.
Uncertainty quantification methods have begun to be used in force field parameterization of simple alkane systems \cite{Messerly2017}; this paper demonstrates the ability to extend the methods to higher-dimensional systems with more diversity of atom types.
Application of these methods offer the benefit of efficiently characterizing parameter space and understanding the impact of parameter variation on quantities of interest.
Additionally, the iterative method we used in the present work is very suitable for this type of problem, as the accuracy of the surrogate models are improved significant after iterations.
Especially, the error of the surrogate models for the atomic charge induced uncertainties are reduced by $40\%\sim 50\%$ compared with the standard compressive sensing method.
Also, there is significant room for development in the numerical methods.
For example, the sparsity-enhancing approaches can be combined with other techniques including improved sampling strategies \cite{RauhutW12, PengHD14}, adaptive basis selection \cite{YangCLK12, JakemanES14}, and advanced optimization methods \cite{CandesWB08, YangK13}.
These approaches improve the accuracy of the compressive sensing method from different aspects.
As such, they will help to reduce the number of expensive simulations or quantum mechanics calculations needed for constructing accurate surrogates.


\section*{Acknowledgments}
This work was supported by the U.S.~Department of Energy, Office of Science, Office of Advanced Scientific-Computing Research as part of the Collaboratory on Mathematics for Mesoscopic Modeling of Materials (CM4) and by NIH grant GM069702.
Pacific Northwest National Laboratory is operated by Battelle for the DOE under Contract DE-AC05-76RL01830.

\bibliographystyle{unsrt}
\bibliography{ref}

%

\end{document}